\documentclass{article}

\usepackage{graphicx}
\newcommand{\ud}{\mathrm{d}}
\begin{document}

\title{Theoretical study of the
${}^{4}\rm{He}(\gamma,p)^3\rm{H}$ and 
${}^{4}\rm{He}(\gamma,n)^3\rm{He}$ reactions
}







\maketitle

\begin{abstract}
{\it Ab initio} calculation of the total
cross section for the reactions   
$^{4}\rm{He}(\gamma,p)^3\rm{H}$ and $^{4}\rm{He}(\gamma,n)^3\rm{He}$
is presented, using state-of-the-art nuclear forces.
The Lorentz integral transform (LIT) method is applied, which allows
exact treatment of the final state interaction (FSI).  
The dynamic equations are solved using the effective interaction
hyperspherical harmonics (EIHH) method.
In this calculation of the cross sections
the three-nucleon force is fully taken into account, 
except in the source term of the LIT equation for the FSI transition matrix element.
\end{abstract}

\section{Introduction}
\label{sec:1_intro}
The photodisintegration of $^{4}\rm{He}$ has been studied extensively. 
Many of the experimental measurements 
are in disagreement with each other for photon energies
$\omega_{\gamma}\!<\!70$ MeV,
see e.g., \cite{Quag04, Raut12, Torn12}.
In the low-energy regime the 
photoabsorption cross section (CS) is dominated by the
$^{4}\rm{He}(\gamma,p)^3\rm{H}$ and $^{4}\rm{He}(\gamma,n)^3\rm{He}$ reactions.
The three- and four-body channels become kinematically accessible at incident
photon energies of 26.1 MeV and 28.3 MeV, respectively. 
However, their contribution is very small. 
Therefore it is not surprising that much of the experimental effort was
directed to the 
$^4\rm{He}(\gamma,p)^3\rm{H}$ and $^4\rm{He}(\gamma,n)^3\rm{He}$ reactions. 
The most recent measurements of these CSs were performed in TUNL \cite{Raut12, Torn12}.
These measurements are in rough agreement with the  
exclusive cross sections calculated in \cite{Quag04} using,
however, a simple semi-realistic nuclear potential model.
In the latter, the Lorentz integral transform (LIT) method \cite{Efro94,Efro07} was applied.
The advent of using the LIT method to calculate the final-state interaction (FSI)
is the ability to obtain the CS of a specific channel regardless of the other channels, 
i.e., even beyond three- and four-body breakup thresholds.
Using other methods, one usually has to account for all the open channels simultaneously.

In the low-energy regime,
the total cross section of the two-body breakup reactions 
is well-described by the unretarded dipole approximation (UDA)
\cite{Elle96,Efro97}.
Therefore meson-exchange currents can be included implicitly using Siegert's theorem.

 Unlike the exclusive case, the total, i.e., inclusive cross section
 of $^{4}\rm{He}$ photoabsorption has already been calculated with state-of-the-art
 nuclear forces, including three-nucleon forces (3NF),
 of either a phenomenological type \cite{Gazi06} or from a chiral perturbation
 effective field theory ($\chi$EFT) \cite{Quag07}.
These two calculations, which also utilize the UDA, are in
excellent agreement with one another, 
and in reasonable agreement with the majority of experimental data.
Therefore, their combined result can serve as a benchmark for
calculations of the exclusive reactions that use modern nuclear models.

A step in that direction was made in \cite{Hori12},
where the exclusive two-body photodisintegration CSs were calculated
using two different nucleon-nucleon interactions that contain central, tensor
and spin-orbit components, supplemented with ad-hoc 3NFs. 
These calculations yield slightly lower CSs than \cite{Quag04}.

In this work we present the first {\it ab-initio}
calculation of the exclusive CSs 
that uses a state-of-the-art nuclear Hamiltonian.
We use the UDA and apply the LIT method to include the FSI. 
The dynamic equations are solved using an extension of the effective interaction hyperspherical harmonics (EIHH)
method that allows: $(i)$ construction of the asymptotic wave functions, 
and $(ii)$ isospin-symmetry breaking.
In this work the 3NF is included in the Hamiltonian and omitted
from the source term of the exclusive LIT equation. 

\section{Theoretical framework}
\label{sec:2_theory}

\subsection{The exclusive cross section}
\label{subsec:CS}
Using partial-waves expansion in the framework of the unretarded dipole
approximation (UDA), the cross section $\sigma_{N,3}$
for photo-induced breakup of $^{4}\rm{He}$ into two fragments,
a nucleon $N$ and an $A=3$ nucleus, is given by 
\begin{eqnarray}
\!\sigma_{(N,3)}\left(\omega_{\gamma}\right)&=&4\pi^{2}\alpha_{D}k\mu\omega_{\gamma}
\sum_{j_N=\frac{1}{2},\frac{3}{2}}
\left|\bigl<\Psi_{J_3,\ell,s_N;j_N,J_T}^{(-)}\left(E_{N,3}\right)
\big|\mathcal{D}_{z}\big|\Psi_{\alpha}\bigr>\right|^{2},\\
E_{N,3}&=&\omega_{\gamma}+E_{\alpha}\,\,\, =\,\,\, T_{rel}+E_3.\nonumber
\label{eq:xclus_CS_with_Dz}
\end{eqnarray}
Here $\alpha_{D}$ is the fine-structure constant, and $\hbar=c=1$ is implied.
$\mu$, $k$ and $T_{rel}=\frac{k^2}{2\mu}$
are, respectively, the reduced mass, relative momentum and kinetic energy of the two fragments,
and $\omega_{\gamma}$ is the photon energy.
$E_{\alpha}$ and $E_3$ are the ground-state energies of the target and
ejectile nuclei.
The continuum wave function
$\Psi_{J_3,\ell,s_N;j_N,J_T}^{(-)}\left(E_{N,3}\right)$
is written in the coupling scheme where the relative angular momentum $\ell$
and the nucleon spin $s_N$ 
are coupled to $j_N$, which is then coupled with the 
mass 3 nuclear spin $J_3$
to a total angular momentum $J_T=1$. 
In the UDA, the transition 
$T_{(N,3)}
=\bigl<\Psi_{J_3,\ell,s_N;j_N,J_T}^{(-)}\left(E_{N,3}\right)
\big|\mathcal{D}_{z}\big|\Psi_{\alpha}\bigr>$
is induced by the dipole operator
$\mathcal{D}_{z}=\sum_{a=1}^4 \frac{1+\tau_a^3}{2} z_a$, 
where $\tau_a^3$ is the third isospin component,
and $z_a$ the position $z$ component of the $a$'th nucleon.

\subsection{The LIT method for exclusive reactions}
\label{subsec:LIT}
The general problem we face here is how to calculate the transition matrix element of the form
\begin{equation}
T_f\left(E_f\right)=\bigl<\Psi_f^{(-)}\left(E_f\right)
\big|\mathcal{O}\big|\Psi_0\bigr>.
\label{eq:T_f}
\end{equation}
Here we will restrict ourselves to the case of two outgoing fragments.
Using the asymptotic function of the outgoing channel $\Phi^-_f\left(E_f\right)$,
the continuum wave function $\Psi_f^{(-)}\left(E_f\right)$
can be written as a solution of the Lippmann-Schwinger equation \cite{Gold64}
\begin{equation}
\big|\Psi_f^{(-)}\left(E_f\right)\big>
=\mathcal{A}\big|\Phi^{(-)}_f\left(E_f\right)\big>
+\left(E_f-i\epsilon-\mathcal{H}\right)^{-1}\mathcal{A}\mathcal{V}\big|\Phi^{(-)}_f\big>, 
\label{eq:Lipp_Scwing}
\end{equation}
where $\mathcal{A}$ is an antisymmetrization operator, and 
$\mathcal{V}$ is the sum of all interactions between particles
belonging to different fragments.\footnote{
In the case where both fragments are charged this is slightly altered, see \cite{Quag04}.}
Plugging Eq.~(\ref{eq:Lipp_Scwing}) into Eq.~(\ref{eq:T_f})
we find that $T_f\left(E_f\right)$ is composed of two terms ---
a Born term, which can be easily calculated, and an FSI term, which constitutes the main difficulty.
Using the LIT method, as described in \cite{Quag04,LaPi00}, the FSI term can
be easily obtained from  
an auxiliary function $F_f\left(E\right)$, whose Lorentz integral transform
\begin{equation}
L\left[F_f\right]\left(\sigma\right)
=\int_{E_{th}^-}^{\infty}
\frac{F_{f}\left(E\right)\ud E}
{\left(E-\sigma_{R}\right)^{2}+\sigma_{I}^{2}},\quad \sigma=\sigma_R+i\sigma_I,\quad \sigma_I>0,
\label{eq:L_of_F}
\end{equation}
can be written as the overlap
$\bigl<\widetilde{\Psi}_f\left(\sigma\right)\big|\widetilde{\Psi}_i\left(\sigma\right)\bigr>$
between the unique solutions of the following Schr\"{o}dinger-like equations
\begin{eqnarray}
\left(\mathcal{H}-\sigma\right)\big|\widetilde{\Psi}_{i}\left(\sigma\right)\big>  & = & \mathcal{O}\big|\Psi_{0}\big>,
\label{eq:LIT_Eq_i} \\
\left(\mathcal{H}-\sigma\right)\big|\widetilde{\Psi}_{f}\left(\sigma\right)\big>
& = & \mathcal{A}\mathcal{V}\big|\Phi_{f}\big>. 
\label{eq:LIT_Eq_f}
\end{eqnarray}
In this work, a simplified version of Eq.~(\ref{eq:LIT_Eq_f}) is used,
where $\mathcal{V}$ does not include the 3NF that appears in $\mathcal{H}$.
As explained in \cite{Quag04}, 
Eqs.~(\ref{eq:LIT_Eq_i}),(\ref{eq:LIT_Eq_f})
can be solved using bound-state methods. 
For this purpose we use a new version of the EIHH method \cite{Barn00,Barn01a,Barn01b,Barn03}
developed in \cite{Nevo08} particularly to solve equations of the form of
Eq.~(\ref{eq:LIT_Eq_f}). 
In addition, differentiating between the two 
channels in the coupled scheme requires a summation over the total isospin of the system $T_A$. 
Therefore, our code was developed with control over the allowed values of
$T_A$, thus enabling either isospin-symmetry breaking (ISB) or 
conservation (ISC). 
As in \cite{Quag04}, the Lanczos technique \cite{Marc03} was used to calculate the overlap 
$\bigl<\widetilde{\Psi}_f\left(\sigma\right)\big|\widetilde{\Psi}_i\left(\sigma\right)\bigr>$  
from which $T_{(N,3)}$ can be 
obtained \cite{Efro07,Efro99}.
In this work, the inversion process required to obtain $F_f\left(E\right)$
is achieved using a combination of two LITs with different values of
$\sigma_I$, as in \cite{Efro10}. 

\section{Results and Discussion}
\label{sec:3_results}
\begin{figure}
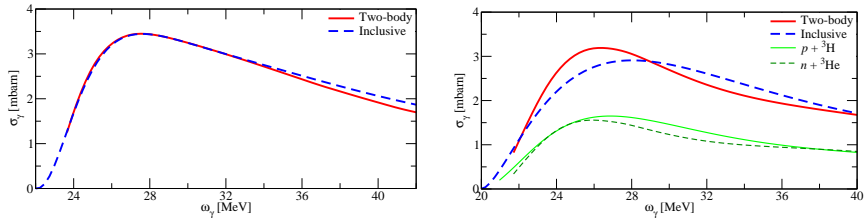

\centering
  \includegraphics[width=0.45\linewidth]{plot.proc_big_font_fig_1_KMT.eps}  
  \hspace{0.02\linewidth}
  \includegraphics[width=0.45\linewidth]{plot.proc_big_font_fig_2_U9AV18_new_color.eps} 
\caption{Colors online.
        (Left) The sum of the two-body $^4\rm{He}$ photodisintegration cross sections
        $\sigma_{n,^3\rm{He}}$ and
        $\sigma_{p,^3\rm{H}}$ (solid curve),
        compared with the inclusive CS (dashed curve);
        all calculated with the KMTI-III NN interaction.
        (Right) The two-body $^4\rm{He}$ photodisintegration cross sections
        $\sigma_{n,^3\rm{He}}$ (thin dashed curve) and
        $\sigma_{p,^3\rm{H}}$ (thin solid curve) and
        their sum (thick solid curve),
        compared with the inclusive CS (thick dashed curve);
        all calculated with the AV18 NN interaction and UIX 3NF.
        }
\label{fig:1_TBCS}
\end{figure}

At first, we tested our ISB mechanism, and found that our calculations 
agree with recent ISB calculations for the ground states of the $A=3,4$ nuclei
\cite{Nogg02,Vivi05,Kiev08}.
Next we turned to the calculation of exclusive $^{4}\rm{He}$ photodisintegration.
To
check our code, we compared the sum of 
$\sigma_{n,^3\rm{He}}$ and $\sigma_{p,^3\rm{H}}$ 
to the inclusive CS obtained from the solution of Eq.~(\ref{eq:LIT_Eq_i})
 as explained, e.g., in \cite{Quag04,Efro94,Efro07,Efro99}.
As a first check, we reproduced the work of \cite{Quag04},
which have used an updated version of the semi-realistic Malfliet-Tjon
nucleon-nucleon (NN) potential~\cite{Malf69}, termed KMTI-III~\cite{Kama92}.
This potential contains no 3NF.
In our calculation we have included only the unretarded
dipole operator, 
which does not induce the $^{4}\rm{He}(\gamma,d)^2\rm{H}$ reaction.
Therefore, the sum $\sigma_{n,^3\rm{He}}+\sigma_{p,^3\rm{H}}$
should coincide with the inclusive CS at least up to the three-body breakup
threshold, 
which lies at $\sim$26.5 MeV for the KMTI-III potential.
The comparison between the sum of the exclusive CSs 
and the inclusive CS is presented in the left panel of
Fig.~\ref{fig:1_TBCS}. 
It can be seen that there is an excellent agreement between the sum of the
exclusive two-body CSs and the inclusive CS.
In this case, the effect of three- and four-body breakup on the CS becomes
visible only above $\sim$34 MeV.

We now turn to study exclusive $^4\rm{He}$ photodisintegration
with realistic state-of-the-art forces.
For the NN interaction we use the AV18 potential \cite{Wiri95}, supplemented
with the UIX 3NF \cite{Pudl95}. 
In the right panel of Fig.~\ref{fig:1_TBCS} we present our results for
$\sigma_{n,^3\rm{He}}$ and $\sigma_{p,^3\rm{H}}$. We also compare their sum
with the inclusive CS.
The large discrepancy between the inclusive CS and the sum of the exclusive
CSs reflects, to our opinion, 
the fact that the 3NF was not included in the r.h.s of Eq.~(\ref{eq:LIT_Eq_f}).
This indicates that the role of the 3NF in the FSI is not negligible despite
its relative weakness and short range. 
Also evident in this figure is the relatively large difference between 
$\sigma_{n,^3\rm{He}}$ and $\sigma_{p,^3\rm{H}}$.
This result can only be obtained when using ISB in the calculation. 
It remains to be seen whether this difference is a pure Coulomb effect, or is
it also affected by charge-symmetry-breaking 
terms in the potential.

\section{Summary}
\label{sec:4_summ}
We have performed an {\it ab initio} calculation of exclusive two-body 
$^4\rm{He}$ photodisintegration CS 
with state-of-the-art nuclear forces.
We have used the LIT method for exclusive reactions,
and the dynamic equations were solved using an updated version of the EIHH method.
From the convergence of the LIT and the variance in the inversion process we
estimate an error of the order of 1\% in the CS. 
For the AV18/UIX potential model we have realized that
the sum of the CSs $\sigma_{n,^3\rm{He}}+\sigma_{p,^3\rm{H}}$ of the reactions
$^{4}\rm{He}(\gamma,p)^3\rm{H}$ and $^{4}\rm{He}(\gamma,n)^3\rm{He}$
does not reproduce the inclusive CS.
This discrepancy is the result of neglecting the 3NF in the source term 
of the exclusive LIT equation~(\ref{eq:LIT_Eq_f}).
We intend to explore this point including also the 3NF in the source term.
The method used here can also be applied for the calculation of 
nucleon-nucleus scattering reactions.
\section{Acknowledgements}
\label{sec:5_acknowledge}
The authors would like to thank Giuseppina Orlandini for useful discussions.
NND would like to thank Sofia Quaglioni for providing her data of the calculation presented in \cite{Quag04}.
This work was supported by the Israel Science Foundation (grant no. 954/09)
and by MIUR grant PRIN-2009TWL3MX.


\begin{thebibliography}{References}
%
%

\bibitem{Quag04}
Quaglioni S.\ \textit{et al.} (2004) %
Phys.\ Rev.\ C \textbf{69}:044002.

\bibitem{Raut12}
Raut R.\ \textit{et al.} (2012) %
Phys.\ Rev.\ Lett.\ \textbf{108}:042502.

\bibitem{Torn12}
Tornow W.\ \textit{et al.} (2012) %
Phys.\ Rev.\ C \textbf{85}:061001(R).

\bibitem{Efro94}
Efros V.\,D.\ \textit{et al.} (1994) 
Phys.\ Lett.\ B \textbf{338}:130--133.

\bibitem{Efro07}
Efros V.\,D.\  \textit{et al.} (2007) 
J.\ Phys.\ G: Nucl.\ Part.\ Phys.\ \textbf{34}:R459--R528.

\bibitem{Elle96}
Ellerkmann G.\ \textit{et al.} (1996) %
Phys.\ Rev.\ C \textbf{53}:2638--2644.

\bibitem{Efro97}
Efros V.\,D.\ \textit{et al.} (1997) %
Phys.\ Rev.\ Lett.\ \textbf{78}:4015--4018.

\bibitem{Gazi06}
Gazit D.\ \textit{et al.} (2006) %
Phys.\ Rev.\ Lett.\ \textbf{96}:112301.

\bibitem{Quag07}
Quaglioni S.\ and Navr\'atil P.\ (2007)
Phys.\ Lett.\ B \textbf{652}:370--375.

\bibitem{Hori12}
Horiuchi W.\ \textit{et al.} (2012)
Phys.\ Rev.\ C \textbf{85}:054002.

\bibitem{Gold64}
Goldberger M.\,L., Watson K.\,W.\ (1964)
        Collision Theory.
Wiley, New York, pg. 197--209.

\bibitem{LaPi00}
La Piana A.\ and Leidemann W.\ (2000)
Nucl.\ Phys.\ \textbf{A677}:423--441.

\bibitem{Barn00}
Barnea N.\ \textit{et al.} (2000) 
Phys.\ Rev.\ C \textbf{61}:054001.

\bibitem{Barn01a}
Barnea N.\ \textit{et al.} (2001) %
Phys.\ Rev.\ C \textbf{63}:057002.

\bibitem{Barn01b}
Barnea N.\ \textit{et al.} (2001) 
Nucl.\ Phys.\ \textbf{A693}:565--578.

\bibitem{Barn03}
Barnea N.\ \textit{et al.} (2003) 
Phys.\ Rev.\ C \textbf{67}:054003.

\bibitem{Nevo08}
Nevo N.\ (2008)
M.Sc.\ thesis, Hebrew University, Jerusalem.

\bibitem{Marc03}
Marchisio M.\,A.\ \textit{et al.} (2003) %
Few-Body Syst.\ \textbf{33}:259--276.

\bibitem{Efro99}
Efros V.\,D.\ \textit{et al.} (1999) 
Few-Body Syst. \textbf{26}:251--269.

\bibitem{Efro10}
Efros V.\,D.\  \textit{et al.} (2010) 
Phys.\ Rev.\ C \textbf{81}:034001.

\bibitem{Nogg02}
Nogga A.\ \textit{et al.} (2002) %
Phys.\ Rev.\ C \textbf{65}:054003.

\bibitem{Vivi05}
Viviani M.\ \textit{et al.} (2005) %
Phys.\ Rev.\ C \textbf{71}:024006.

\bibitem{Kiev08}
Kievsky A.\ \textit{et al.} (2008) %
J.\ Phys.\ G: Nucl.\ Part.\ Phys.\ \textbf{35}:063101.

\bibitem{Malf69}
Malfliet R.\,A.\ and Tjon J.\ (1969)
Nucl.\ Phys.\ \textbf{A127}:161--168.

\bibitem{Kama92}
Kamada H.\  and Gl\"ockle W.\ (1992)
Nucl.\ Phys.\ \textbf{A548}:205--226.

\bibitem{Wiri95}
Wiringa R.\,B.\ \textit{et al.} (1995) 
Phys.\ Rev.\ C \textbf{51}:38--51.

\bibitem{Pudl95}
Pudliner B.\,S.\ \textit{et al.} (1995) %
Phys.\ Rev.\ Lett.\ \textbf{74}:4396--4399.
\end{thebibliography}
%

\end{document}